\documentclass[a4paper]{article}
\usepackage{geometry}
\usepackage{amsmath}
\usepackage{amsfonts}
\usepackage{amsmath,amssymb}
\usepackage{graphicx}
\usepackage{color}

\newtheorem{theorem}{Theorem}
\newtheorem{lemma}{Lemma}

\newtheorem{definition}{Definition}

\newtheorem{remark}{Remark}
\newtheorem{conjecture}{Conjecture}

\newcommand{\ls}[1]
    {\dimen0=\fontdimen6\the\font\lineskip=#1\dimen0
     \advance\lineskip.5\fontdimen5\the\font
     \advance\lineskip-\dimen0
     \lineskiplimit=0.9\lineskip
     \baselineskip=\lineskip
     \advance\baselineskip\dimen0
     \normallineskip\lineskip\normallineskiplimit\lineskiplimit
     \normalbaselineskip\baselineskip
     \ignorespaces}

\begin{document}

\bibliographystyle{abbrv}

\title{Autocorrelation and Lower Bound on the 2-Adic Complexity of LSB Sequence of $p$-Ary $m$-Sequence}
\author{Yuhua Sun$^{\mathrm{a}}$, Qiuyan Wang$^{\mathrm{b}}$, Tongjiang Yan$^{\mathrm{a}}$, Chun'e Zhao$^{\mathrm{a,c}}$\\
$^a$ College of Science,
China University of Petroleum,\\
Qingdao 266555,
Shandong, China\\
$^b$ School of Computer Science and Technology, \\
Tiangong University, Tianjin 300387,China)\\
$^c$ Key Laboratory of Applied Mathematics(Putian University),\\
Fujian Province University, Fujian Putian, 351100, China.\\
Corresponding author: Qiuyan Wang\\
Email: sunyuhua\_1@163.com; wangyan198801@163.com;\\
 yantoji@163.com; zhaochune1981@163.com\\
}
\maketitle
\thispagestyle{plain} \setcounter{page}{1}
\begin{abstract}
LSB (Least Significant Bit) sequences are widely used as the initial inputs in some modern stream ciphers, such as the ZUC algorithm-the core of the 3GPP LTE International Encryption Standard. Therefore, analyzing the statistical properties
(for example, autocorrelation, linear complexity and 2-adic complexity) of these sequences becomes an important research topic.
In this paper, we first reduce the autocorrelation distribution of the LSB  sequence of a $p$-ary $m$-sequence with period $p^n-1$ for any order $n\geq2$ to the autocorrelation distribution of a corresponding Costas sequence with period $p-1$, and from the computing of which by computer, we obtain the explicit autocorrelation distribution of the LSB sequence for each prime $p<100$. In addition, we give  a lower bound on the 2-adic complexity of each of these LSB sequences for all primes $p < 20$, which proves to be large enough to resist the analysis of RAA (Rational Approximation Algorithm) for FCSRs (Feedback with Carry Shift Registers). In particular, for
a Mersenne prime $p=2^k-1$ (i.e., $k$ is a prime such that $p$ is also a prime), our results hold for all its bit-component sequences  since they are shift equivalent to the LSB sequence.



\end{abstract}

\ls{1.2}
\section{Introduction}\label{section 1}
Pseudo-random sequences with good correlation and large linear complexity have widely applications in communication systems and cryptography. Due to their ideal correlation property and other good performance measures such as  highly efficient implementation, maximal length linear feedback shift register (LFSR) sequences (i.e., $m$-sequences) have been widely used in designing stream ciphers. However, since the linear complexity of these sequences is relatively low under the analysis of Berlekamp-Massey Algorithm (BMA), they can not be used by themselves. Therefore constructing nonlinear sequence generators with desirable good properties become a very important topic. As one class of promising nonlinear sequence generators, feedback with carry shift registers (FCSRs), were originally presented by Klapper and Goresky in 1997, and the  notion of 2-adic complexity $\Phi_{2}(s)$ for a binary periodic sequence $s$, i.e., the length of the shortest FCSR which generates $s$, was introduced in \cite{Andrew Klapper}. One direct result of this notion is that an $m$-sequence with period $N=2^n-1$ has maximal 2-adic complexity if $2^N-1$ is a prime. Similar to BMA of LFSRs, Klapper and Goresky also proposed an algorithm, called Rational Approximation Algorithm (RAA), to determine the 2-adic complexity of $s$ and showed that, from the perspective of cryptography security,  a desirable sequence should has both high linear complexity and high 2-adic complexity, namely, greater than or equal to one half of the period. Although the linear complexity of many classes of sequences have been obtained (see \cite{Ding Cunsheng,Edemskiy,Etzion,Helleseth,Hu Liqin,Agnes Hui,Yong,Li Nian,Rainer,Qi Wang,Wang Qiu,Xiong}), there are only a
handful of papers on their 2-adic complexity. After Tian and Qi made a breakthrough, i.e., they proved that all binary $m$-sequences have maximal 2-adic complexity in \cite{Tian Tian},
Xiong et al. presented a new method to compute the 2-adic complexity of binary sequences by circulant matrixes in \cite{Xiong Hai, Xiong Hai-2}. They showed that all the known
sequences with ideal 2-level autocorrelation and several other classes of sequences with optimal autocorrelation have maximum 2-adic complexity. Then Hu presented a simpler method in \cite{Hu Honggang} to obtain the results of Xiong et al. by using exact autocorrelation distributions. More applications of these two methods can be
found in \cite{Zeng Xiangyong,Hofer,Sun Yuhua-1,Sun Yuhua-2,Sun Yuhua-3}, in which the 2-adic complexity of
 Legendre sequences, Jacobi sequences and modified Jacobi sequences was analyzed.

Since LSB sequences of $p$-ary $m$-sequences (see Definitions \ref{defofbitcomponent-thresh-1}) can be easily implemented and have been tested to possess many good pseudo-random properties, some modern stream ciphers, such as the ZUC algorithm-the core of the 3GPP LTE International Encryption Standard, are designed by using them as the inputs \cite{3GPP-1,3GPP-2}. Earlier, Chan and Games \cite{Agnes Hui} proved that these sequences have high linear complexity. However, the autocorrelation and the 2-adic complexity of them have still not been studied as far as we know.

The rest of this paper is organized as follows. We introduce notations and some well-known results in Section 2. Some autocorrelation
properties of LSB sequences of $p$-ary $m$-sequences, as well as the explicit autocorrelation distributions of
 Costas sequences with period $p-1$ for $p<100$, are given in Section 3. In Section 4, the lower bound on the 2-adic complexity
of each of the LSB sequences of $p$-ary $m$-sequences for $p<20$ and an open problem on the 2-adic complexity of the LSB sequence of a $p$-ary $m$-sequence for any prime $p$ are presented.
\section{Preliminaries}\label{section 2}
Let $N$ be a positive integer and $s=(s_{0},s_{1},\cdots,s_{N-1},\cdots)$ a binary sequence of period $N$. The autocorrelation of $s$ is given by
\begin{equation}
AC_{s}(\tau)=\sum_{t=0}^{N-1}(-1)^{s_t+s_{t+\tau}},\ \tau=0,1,2,\cdots,N-1.\label{AC}
\end{equation}
Let $S(x)=\sum\limits_{i=0}^{N-1}s_{i}x^i\in \mathbb{Z}[x]$. Then we write
\begin{equation}
\frac{S(2)}{2^N-1}=\frac{\sum\limits_{i=0}^{N-1}s_{i}2^i}{2^N-1}=\frac{C}{D},\ 0\leq C\leq D,\ \mathrm{gcd}(C,D)=1,\label{2-adic complexity}
\end{equation}
where $\mathrm{gcd}(x,y)$ is the greatest common divisor of $x$ and $y$.
The 2-adic complexity $\Phi_{2}(s)$ of the sequence $s$ is the integer $\lfloor\mathrm{log}_2 D\rfloor$, i.e.,
\begin{equation}
\Phi_{2}(s)=\left\lfloor\mathrm{log}_2\frac{2^N-1}{\mathrm{gcd}(2^N-1,S(2))}\right\rfloor,\label{2-adic calculation}
\end{equation}
where $\lfloor x\rfloor$ is the greatest integer that is less than or equal to $x$.

Let $p$ be any odd prime, $n$ a positive integer, and $\alpha$ a primitive element of $\mathbb{F}_{p^n}$. Then
\begin{equation}
a_t=\mathrm{Tr}(\alpha^t),\ t=0,1,2,\cdots,p^n-2,\label{m-definition}
\end{equation}
is a $p$-ary $m$-sequence, where $\mathrm{Tr}(x)=x+x^p+x^{p^2}+\cdots+x^{p^{n-1}}$ is the trace function from $\mathbb{F}_{p^n}$ to $\mathbb{F}_{p}$.

For each term $a_{t}$ of the $m$-sequence $\{a_t\}_{t=0}^{p^n-2}$, we have the following 2-adic expansion
$$
a_t=a_{t,0}+a_{t,1}\times2+a_{t,2}\times2^2+\cdots+a_{t,k-1}\times2^{k-1},\ a_{t,i}\in\{0,1\},\ i=0,1,\cdots,k-1,
$$
where $k=\lceil \mathrm{log}_2p\rceil$ and $\lceil x\rceil$ is the least integer that is larger than or equal to $x$.
Here, we identify the bit string $(a_{t,0},a_{t,1},\cdots,a_{t,k-1})$ of length $k$ with the element $a_t$ and the $i$-th element $a_{t,i-1}$ is called as the $i$-th bit-component of $a_t$. But the element $0\in \mathbb{F}_{p}$ is written as $p$, i.e., 0 is identified with  $(p_0,p_1,\cdots,p_{k-1})$, where the 2-adic expansion of $p$ is $p_0+p_1\times2+\cdots+p_{k-1}\times2^{k-1}$ (this is in accordance with the ZUC algorithm).
\begin{definition}\label{defofbitcomponent-thresh-1}
For a fixed $i\in\{1,2,\cdots,k\}$, the sequence
$\{a_{t,i-1}\}_{t=0}^{p^n-2}\label{bit-componen-seq}$
 is called the $i$-th bit-component sequence of $\{a_t\}_{t=0}^{p^n-2}$. In particular, when $i=0$,  the bit-component sequence $\{a_{t,0}\}_{t=0}^{p^n-2}$ is called the Least Significant Bit sequence (the LSB sequence) of the $m$-sequence $\{a_t\}_{t=0}^{p^n-2}$ and we denote $\{s_t\}_{t=0}^{p^n-2}=\{a_{t,0}\}_{t=0}^{p^n-2}$ for convenience. In fact, it can also be expressed as
\begin{equation}\label{defofLSB}
 s_{t}=\left\{
 \begin{array}{ll}
 \mathrm{Tr}(\alpha^t)\ (\mathrm{mod}\ 2),\ \ \mathrm{if\ Tr}(\alpha^t)\in \mathbb{F}_p^{\ast},\\
 1,\ \ \ \ \ \ \ \ \ \ \ \ \ \ \ \ \ \ \ \mathrm{if\ Tr}(\alpha^t)=0.
 \end{array}
 \right.
\end{equation}
\end{definition}

\begin{definition}\label{defofbitcomponent-thresh-2}
Let $p$ be any odd prime, $n$ a positive integer, and $\alpha$ a primitive element of $\mathbb{F}_{p^n}$.
Let $\beta=\alpha^{\frac{p^n-1}{p-1}}$, a primitive element of $\mathbb{F}_p$.
The Costas sequence is  defined as the sequence $\{b_j\}_{j=0}^{p-2}$ of period $p-1$ which is given by
$
b_j\equiv\beta^{j}\ (\mathrm{mod}\ 2).
$
\end{definition}

The Costas sequence  $\{b_j\}_{j=0}^{p-2}$ is actually the LSB sequence of  the permutation $\{\beta^0, \beta^1, \ldots, \beta^{p-2} \}$ corresponding to a Welch Costas array determined by the primitive element $\beta$.
This sequence was first considered by J. P Costas  in 1984 as permutation matrices with ambiguity functions taking only the values 0 and (possibly) 1, applied to the processing of radar and sonar signals. The basic algebraic construction of this sequence can be found in \cite{Golomb}. The sequence is closely related to APN functions and S-Box of block ciphers \cite{Konstantinos}.
\begin{definition} A function from $\mathbb{F}_{p^{n}}$ to $\mathbb{F}_{p}$ is said to be balanced if the element 0 appears one less time than each nonzero element in $\mathbb{F}_{p}$ in the list $f(\alpha^{0}),\ f(\alpha^{1}),\cdots,f(\alpha^{p^{n}-2})$, where $\alpha$ is a primitive element of $\mathbb{F}_{p^{n}}$.
\end{definition}
\begin{definition}\label{D-B-F} Let $f(x)$ be a function on $\mathbb{F}_{p^{n}}$ over $\mathbb{F}_{p}$. Then the function $f(x)$ is called difference-balanced if $f(xz)-f(x)$ is balanced for any $z\in \mathbb{F}_{p^{n}}$ but $z\neq1$.
\end{definition}
\begin{remark}\label{rem-2}
It is well known that the trace function $\mathrm{Tr}(x)$ from $\mathbb{F}_{p^n}$ to $\mathbb{F}_p$ is difference-balanced, which is in fact a linear function over $\mathbb{F}_p$.
\end{remark}

\section{Autocorrelation properties of LSB sequences of $p$-ary $m$-sequences}\label{section 3}
For the rest of the paper, we denote $N=p^n-1$, $M=\frac{N}{p-1}$,  and $\mathbb{Z}_N=\{0,1,2,\cdots,N-1\}$ unless specified.

\begin{lemma}\label{trival proper}
Let $n\geq2$. Then, for $0<\tau<N$ and $\tau\notin\{M\tau^{\prime}\mid\tau^{\prime}=1,2,\cdots,p-2\}$, the autocorrelation value $AC_{s}(\tau)$ of $\{s_{t}\}_{t=0}^{N-1}$ is given by
$AC_{s}(\tau)=p^{n-2}-1.$
\end{lemma}
$\mathbf{Proof.}$
For a fixed $\tau$, we denote $D_{\tau}=\{t\mid s_t\neq s_{t+\tau},\ t\in \mathbb{Z}_N\}$. Then we get
\begin{equation}
AC_{s}(\tau)=|\mathbb{Z}_N\setminus D_{\tau}|-|D_{\tau}|=N-2|D_{\tau}|,\label{(1)}
\end{equation}
where $|D_{\tau}|$ is the number of the elements in $D_{\tau}$.
By Eq. (\ref{defofLSB}) in Definition \ref{defofbitcomponent-thresh-1}, we know
\begin{eqnarray}
&&|D_{\tau}|=|\{t\mid s_t\neq s_{t+\tau},\ t\in \mathbb{Z}_N\}|\nonumber\\
&=&|\{t  \in \mathbb{Z}_N \mid\mathrm{Tr}(\alpha^{t}), \mathrm{Tr}(\alpha^{t+\tau})\in \mathbb{F}_p^{\ast},\ \mathrm{Tr}(\alpha^{t})\equiv1\ (\mathrm{mod}\ 2), \mathrm{Tr}(\alpha^{t+\tau})\equiv0\ (\mathrm{mod}\ 2)\}|\nonumber\\
&+&|\{t \in \mathbb{Z}_N\mid\mathrm{Tr}(\alpha^{t}), \mathrm{Tr}(\alpha^{t+\tau})\in \mathbb{F}_p^{\ast},\ \mathrm{Tr}(\alpha^{t})\equiv0\ (\mathrm{mod}\ 2),  \mathrm{Tr}(\alpha^{t+\tau})\equiv1\ (\mathrm{mod}\ 2) \}|\nonumber\\
&+&|\{t\mid\mathrm{Tr}(\alpha^{t})=0,\ \mathrm{Tr}(\alpha^{t+\tau})\in \mathbb{F}_p^{\ast}\ \mathrm{and}\ \mathrm{Tr}(\alpha^{t+\tau})= 0\ (\mathrm{mod}\ 2),\ t\in \mathbb{Z}_N\}|\nonumber\\
&+&|\{t\mid\mathrm{Tr}(\alpha^{t+\tau})=0,\ \mathrm{Tr}(\alpha^{t})\in \mathbb{F}_p^{\ast}\ \mathrm{and}\ \mathrm{Tr}(\alpha^{t})= 0\ (\mathrm{mod}\ 2),\ t\in \mathbb{Z}_N\}|\nonumber\\
&=&|\{x  \in \mathbb{F}_{p^n}^{\ast}  \mid\mathrm{Tr}(x), \mathrm{Tr}(\alpha^{\tau}x)\in \mathbb{F}_p^{\ast},\ \mathrm{Tr}(x)\equiv1\ (\mathrm{mod}\ 2),  \mathrm{Tr}(\alpha^{\tau}x)\equiv0\ (\mathrm{mod}\ 2)\}|\label{1-1}\\
&+&|\{x  \in \mathbb{F}_{p^n}^{\ast}  \mid\mathrm{Tr}(x), \mathrm{Tr}(\alpha^{\tau}x)\in \mathbb{F}_p^{\ast},\ \mathrm{Tr}(x)\equiv0\ (\mathrm{mod}\ 2),  \mathrm{Tr}(\alpha^{\tau}x)\equiv1\ (\mathrm{mod}\ 2) \}|\label{1-2}\\
&+&|\{x\in \mathbb{F}_{p^n}^{\ast} \mid\mathrm{Tr}(x)=0,\ \mathrm{Tr}(\alpha^{\tau}x)\in \mathbb{F}_p^{\ast}, \mathrm{and}\  \mathrm{Tr}(\alpha^{\tau}x)= 0\ (\mathrm{mod}\ 2)\}|\label{1-3}\\
&+&|\{x\in \mathbb{F}_{p^n}^{\ast}\mid\mathrm{Tr}(\alpha^{\tau}x)=0,\ \mathrm{Tr}(x)\in \mathbb{F}_p^{\ast},\mathrm{and}\ \mathrm{Tr}(x)= 0\ (\mathrm{mod}\ 2) \}|.\label{1-4}
\end{eqnarray}
Next, we determine the values of expressions (\ref{1-1})-(\ref{1-4}) respectively.
From Definition \ref{defofbitcomponent-thresh-1}, it is obvious that
$$s_t\neq s_{t+\tau}\Rightarrow\mathrm{Tr}(\alpha^t)-\mathrm{Tr}(\alpha^{t+\tau})\neq0\Rightarrow\mathrm{Tr}(x)-\mathrm{Tr}(\alpha^{\tau}x)\neq0,\ \mathrm{where}\ x=\alpha^t.
$$
By Remark \ref{rem-2} we know that the trace function $\mathrm{Tr}(x)$ is difference-balanced, namely, for each fixed $a\in \mathbb{F}_p^{\ast}$, the total number of $x$'s in $\mathbb{F}_{p^n}^{\ast}$ satisfying the equation $\mathrm{Tr}(x)-\mathrm{Tr}(\alpha^\tau x)=a$
is $p^{n-1}$. And the number of $x$'s to the equation $\mathrm{Tr}(x)-\mathrm{Tr}(\alpha^\tau x)=a$ is actually the sum of the numbers of solutions $x$'s to the following system of equations
\begin{equation}\label{equaoftra}
\left\{
\begin{array}{ll}
\mathrm{Tr}(x)=c+a,\\
\mathrm{Tr}(\alpha^{\tau}x)=c,
\end{array}
\right.
\end{equation}
where $c$ runs through $\mathbb{F}_{p}$.
Notice that $\mathbb{F}_{p^n}$ is an $n$-dimensional vector space over $\mathbb{F}_{p}$. Let $\{\alpha_{1},\alpha_{2},\cdots,\alpha_{n}\}$ be a basis of $\mathbb{F}_{p^n}$ over $\mathbb{F}_{p}$. For any element $x\in \mathbb{F}_{p^n}$, there exist $n$ elements $x_{i}\in \mathbb{F}_{p},\ i=1,2,\cdots,n,$ such that $x=\sum\limits_{i=1}^{n}x_{i}\alpha_{i}$. Then, for fixed $c+a\in \mathbb{F}_{p}$, $c\in \mathbb{F}_{p}$ and $\alpha^{\tau}\in\mathbb{F}_{p^n}$, Eq. (\ref{equaoftra}) can be transformed into
\begin{equation}\label{equaoftra-1}
\left\{
\begin{array}{ll}
\sum\limits_{i=1}^{n}\mathrm{Tr}(\alpha_{i})x_{i}=c+a,\\
\sum\limits_{i=1}^{n}\mathrm{Tr}(\alpha^{\tau}\alpha_{i})x_{i}=c,
\end{array}
\right.
\end{equation}
which is a linear equation system over $\mathbb{F}_{p}$ with $n$ unknowns $x_{i}\in \mathbb{F}_{p},\ i=1,2,\cdots,n,$ and its coefficient matrix is
\begin{equation}\label{equaoftra-2}
A=\left(
\begin{array}{cccc}
\mathrm{Tr}(\alpha_{1})&\mathrm{Tr}(\alpha_{2})& \cdots & \mathrm{Tr}(\alpha_{n})\\
\mathrm{Tr}(\alpha^{\tau}\alpha_{1})&\mathrm{Tr}(\alpha^{\tau}\alpha_{2})& \cdots & \mathrm{Tr}(\alpha^{\tau}\alpha_{n})
\end{array}
\right).
\end{equation}
In fact, for $\alpha^{\tau}\notin \mathbb{F}_{p}^{\ast}$, i.e., $\tau\notin\{M\tau^{\prime}\mid\tau^{\prime}=1,2,\cdots,p-1\}$, the two rows in the above matrix $A$ are linearly independent. Otherwise, there is an element $\delta\in\mathbb{F}_{p}$ such that $\mathrm{Tr}(\alpha^{\tau}\alpha_{i})=\delta\mathrm{Tr}(\alpha_{i})$ for each $i\in\{1,2,\cdots,n\}$, i.e., $\mathrm{Tr}((\alpha^{\tau}-\delta)\alpha_{i})=0$ for each $i\in\{1,2,\cdots,n\}$, which results in $\mathrm{Tr}((\alpha^{\tau}-\delta)\gamma)$ for each element $\gamma\in\mathbb{F}_{p^n}$ since $\{\alpha_{1},\alpha_{2},\cdots,\alpha_{n}\}$ is a basis of $\mathbb{F}_{p^n}$ over $\mathbb{F}_{p}$. This is impossible since $\alpha^{\tau}-\delta\neq0$. Therefore, the rank of the above matrix $A$ in Eq.(\ref{equaoftra-2}) is 2, which implies that there are $p^{n-2}$ solutions in $\mathbb{F}_{p^n}$ to the Equation System (\ref{equaoftra}) for each $a\in \mathbb{F}_{p}^{\ast}$ and $c\in \mathbb{F}_{p}$.

Note that there are exactly $\frac{p-1}{2}$ $c$'s in $\mathbb{F}_p^{\ast}$ such that $c\equiv0\ (\mathrm{mod}\ 2)$ and there are $\frac{p-1}{2}$ $a$'s in $\mathbb{F}_p^{\ast}$ such that $c+a\equiv1\ (\mathrm{mod}\ 2)$ for each fixed $c\equiv0\ (\mathrm{mod}\ 2)$ in $\mathbb{F}_p^{\ast}$. Then the value of Expression (\ref{1-1}) is equal to
\begin{eqnarray}
|\{x\in \mathbb{F}_{p^n}^{\ast}\mid\mathrm{Tr}(x), \mathrm{Tr}(\alpha^{\tau}x)\in \mathbb{F}_p^{\ast},\ \mathrm{Tr}(x)\equiv1\ (\mathrm{mod}\ 2), \mathrm{Tr}(\alpha^{\tau}x)\equiv0\ (\mathrm{mod}\ 2)\}|=p^{n-2}\times\frac{(p-1)^2}{4}.\label{sum-1}
\end{eqnarray}
Similarly, we obtain 
\begin{eqnarray}
|\{x\in \mathbb{F}_{p^n}^{\ast} \mid\mathrm{Tr}(x), \mathrm{Tr}(\alpha^{\tau}x)\in \mathbb{F}_p^{\ast},\ \mathrm{Tr}(x)\equiv0\ (\mathrm{mod}\ 2), \mathrm{Tr}(\alpha^{\tau}x)\equiv1\ (\mathrm{mod}\ 2) \}|
=p^{n-2}\times\frac{(p-1)^2}{4},\label{sum-2}
\end{eqnarray}
and
\begin{eqnarray}
&&|\{x\in \mathbb{F}_{p^n}^{\ast}  \mid\mathrm{Tr}(x)=0,\ \mathrm{Tr}(\alpha^{\tau}x)\in \mathbb{F}_p^{\ast},\ \mathrm{and}\ \mathrm{Tr}(\alpha^{\tau}x)=0\ (\mathrm{mod}\ 2)\}|
=p^{n-2}\times\frac{p-1}{2}, \label{sum-3}\\
&&|\{x\in \mathbb{F}_{p^n}^{\ast}  \mid\mathrm{Tr}(\alpha^{\tau}x)=0,\ \mathrm{Tr}(x)\in \mathbb{F}_p^{\ast},\ \mathrm{and}\ \mathrm{Tr}(x)= 0\ (\mathrm{mod}\ 2)\}|
=p^{n-2}\times\frac{p-1}{2}\label{sum-4}
\end{eqnarray}
respectively. Using the above Eqs. (\ref{sum-1})-(\ref{sum-4}), we get
$$
|D_{\tau}|=p^{n-2}\left(\frac{(p-1)^2}{4}+\frac{(p-1)^2}{4}+\frac{p-1}{2}+\frac{p-1}{2}\right)=\frac{p^n-p^{n-2}}{2}.
$$
By Eq. (\ref{(1)}), $AC_{s}(\tau)=N-2\times\frac{p^n-p^{n-2}}{2}=p^{n}-1-(p^n-p^{n-2})=p^{n-2}-1.$ \ \ \ \ \ \ \ \ \ \ \ \ \ \ \ \ \ \ \ \ \ \ \ \ \ \ \ \ \ \ \ $\blacksquare$

\begin{lemma}\label{nontrivil pro}
For $\tau\in\{M\tau^{\prime}\mid\tau^{\prime}=1,2,\cdots,p-2\}$, the autocorrelation of the LSB sequence $\{s_t\}_{t=0}^{p^n-1}$ satisfies the relation
$AC_{s}(\tau)=\left(AC_b(\tau^{\prime})+1\right)p^{n-1}-1,$
where the sequence $\{b_j\}_{j=0}^{p-2}$ is defined as in Definition \ref{defofbitcomponent-thresh-2}.
\end{lemma}
$\mathbf{Proof.}$ Recall that $\alpha^{\tau}=\beta^{\tau^{\prime}}\in \mathbb{F}_p^{\ast}$ for $\tau\in\{M\tau^{\prime}\mid\tau^{\prime}=1,2,\cdots,p-2\}$
since $\beta=\alpha^M$ and $M=\frac{p^n-1}{p-1}$. Then $\mathrm{Tr}(\alpha^{\tau}x)=\mathrm{Tr}(\beta^{\tau^{\prime}}x)=\beta^{\tau^{\prime}}\mathrm{Tr}(x)$ for $x\in \mathbb{F}_{p^n}^{\ast}$,i.e.,
$
\mathrm{Tr}(\alpha^{\tau}x)\in \mathbb{F}_p^{\ast}\Leftrightarrow \mathrm{Tr}(x)\in \mathbb{F}_p^{\ast}.\label{condition}
$
It is similar to the proof of Lemma \ref{trival proper}, $AC_{s}(\tau)=|\mathbb{Z}_N\setminus D_{\tau}|-|D_{\tau}|=N-2|D_{\tau}|$, where $D_{\tau}=\{t\mid s_{t}\neq s_{t+\tau},\ t\in \mathbb{Z}_N\}$, and
\begin{eqnarray}
s_t\neq s_{t+\tau}&\Rightarrow\mathrm{Tr}(x)-\mathrm{Tr}(\alpha^{\tau}x)\neq0\Rightarrow\mathrm{Tr}(x)\in \mathbb{F}_p^{\ast}\ \mathrm{and}\ \mathrm{Tr}(\alpha^{\tau}x)\in \mathbb{F}_p^{\ast},\ \mathrm{where}\ x=\alpha^t.\label{conditon-2}
\end{eqnarray}
Therefore,
\begin{eqnarray}
|D_{\tau}|&=&|\{x\mid\mathrm{Tr}(x)\in \mathbb{F}_p^{\ast},\ \mathrm{Tr}(x)\not\equiv\beta^{\tau^{\prime}}\mathrm{Tr}(x)\ (\mathrm{mod}\ 2),\ x\in \mathbb{F}_{p^n}^{\ast}\}|\nonumber\\
&=&p^{n-1}\times|\{(c,\beta^{\tau^{\prime}}c)\mid c\in\mathbb{F}_p^{\ast},\ c\not\equiv\beta^{\tau^{\prime}}c\ (\mathrm{mod}\ 2)\}|\label{2-1-1}\\
&=&p^{n-1}\times|\{j\mid\beta^{j}\not\equiv\beta^{j+\tau^{\prime}}\ (\mathrm{mod}\ 2),\ j=0,1,\cdots,p-2\}|\nonumber\\
&=&p^{n-1}\times|D_{\tau^{\prime}}^{\prime}|,\label{2-1}
\end{eqnarray}
where $D_{\tau^{\prime}}^{\prime}=\{j\mid\beta^{j}\not\equiv\beta^{j+\tau^{\prime}}\ (\mathrm{mod}\ 2),\ j=0,1,\cdots,p-2\}$ and Eq. (\ref{2-1-1}) holds because the equation $\mathrm{Tr}(x)=c$ has exact $p^{n-1}$ solutions in $\mathbb{F}_{p^n}^{\ast}$ for each fixed $c\in \mathbb{F}_p^{\ast}$. Hence, we have
\begin{eqnarray}
AC_s(\tau)=N-2|D_{\tau}|=(p^n-1)-2p^{n-1}|D_{\tau^{\prime}}^{\prime}|=(p-2 |D_{\tau^{\prime}}^{\prime}|)p^{n-1}-1.\label{relation}
\end{eqnarray}
Furthermore, since the autocorrelation of the Costas sequence $\{b_j\}_{j=0}^{p-2}$ is equal to
\begin{equation}AC_b(\tau^{\prime})=\sum\limits_{j=0}^{p-2}(-1)^{b_j-b_{j+\tau^{\prime}}}=|\mathbb{Z}_{p-1}\setminus D_{\tau^{\prime}}^{\prime}|-|D_{\tau^{\prime}}^{\prime}|=p-1-2|D_{\tau^{\prime}}^{\prime}|,\label{auto- of-b}
\end{equation}
the result follows.\ \ \ \ \ \ \ \ \ \ \ \ \ \ \ \ \ \ \ \ \ \ \ \ \ \ \ \ \ \ \ \ \ \ \ \ \ \ \ \ \ \ \ \ \ \ \ \ \ \ \ \ \ \ \ \ \ \ \ \ \ \ \ \ \ \ \ \ \ \ \ \ \ \ \ \ \ \ \ \ \ \ \ \ \ \ \ \ \ \ \ \ \ \ \ \ \ \ \ \ \ \ \ \ \ \ \ \ \ \ \ \ \ $\blacksquare$

Combining the results of Lemmas \ref{trival proper} and \ref{nontrivil pro}, we have simplified the problem of computing the autocorrelation of the LSB sequence $\{s_t\}_{t=0}^{N-1}$ of period $p^n-1$ for any positive integer $n\geq2$ to the problem of computing the autocorrelation of the Costas sequence $\{b_j\}_{j=0}^{p-2}$ of period $p-1$.

\begin{lemma}\label{nontri prop}
Let the symbols be the same as above. We have the following results.
\begin{itemize}
\item[(1)] For $1\leq \tau^{\prime}\leq \frac{p-3}{2}$, $AC_{b}(p-1-\tau^{\prime})=AC_{b}(\tau^{\prime}).$
\item[(2)] For $p\equiv1\pmod{4}$ and $1\leq \tau^{\prime}\leq \frac{p-1}{4}$ or for $p\equiv3\pmod{4}$ and $1\leq \tau^{\prime}\leq \frac{p-3}{4}$, $AC_{b}(\frac{p-1}{2}-\tau^{\prime})=-AC_{b}(\tau^{\prime})$. Particularly, when $p\equiv1\pmod{4}$,  $AC_{b}(\frac{p-1}{4})=0$.
\item[(3)] $AC_{b}(\frac{p-1}{2})=-(p-1)$.
\end{itemize}
\end{lemma}
$\mathbf{Proof:}$ (1) By the discussion in Lemma \ref{nontrivil pro}, for a fixed $1\leq \tau^{\prime}\leq p-2$, the autocorrelation value $AC_{b}(\tau^{\prime})$ depends on $|D_{\tau^{\prime}}|$ which is in fact the number of $c$'s in $\mathbb{F}_{p}^{\ast}$ such that the pair $(c,\beta^{\tau^{\prime}}c)$ has different least significant bit (See Eqs. (\ref{2-1-1})-(\ref{2-1})).
Let $c^{\prime}=\beta^{\tau^{\prime}}c$ for $1\leq \tau^{\prime}\leq \frac{p-3}{2}$. Then $(c,\beta^{\tau^{\prime}}c)=(\beta^{-\tau^{\prime}} c^{\prime},c^{\prime})=(\beta^{p-1-\tau^{\prime}}c^{\prime},c^{\prime})$. Since $c^{\prime}$ runs exactly through $\mathbb{F}_p^{\ast}$ when $c$ runs through $\mathbb{F}_p^{\ast}$,
we have $|D^{\prime}_{\tau^{\prime}}|=|D^{\prime}_{p-1-\tau^{\prime}}|$, which implies $AC_{b}(p-1-\tau^{\prime})=AC_{b}(\tau^{\prime})$ by Eq. (\ref{auto- of-b}). \\
(2) Since $-c$ is odd if $c$ is even for $c\in \mathbb{F}_p^{\ast}$ and vice versa (Notice that $p$ is odd and $-c=p-c$), we can derive
$|\{(-c,\beta^{\tau^{\prime}}c)\mid c\in\mathbb{F}_p^{\ast},\ -c\not\equiv\beta^{\tau^{\prime}}c\ (\mathrm{mod}\ 2)\}|=(p-1)-|\{(c,\beta^{\tau^{\prime}}c)\mid c\in\mathbb{F}_p^{\ast},\ c\not\equiv\beta^{\tau^{\prime}}c\ (\mathrm{mod}\ 2)\}|$,
which results in $-AC_{b}(\tau^{\prime})=(p-1)-|\{(-c,\beta^{\tau^{\prime}}c)\mid c\in\mathbb{F}_p^{\ast},\ -c\not\equiv\beta^{\tau^{\prime}}c\ (\mathrm{mod}\ 2)\}|$.
Let $c^{\prime}=\beta^{\tau^{\prime}}c$. Then $(-c,\beta^{\tau^{\prime}}c)=(\beta^{\frac{p-1}{2}-\tau^{\prime}}c^{\prime},c^{\prime})$ from $\beta^{\frac{p-1}{2}}=-1$.
By Eq. (\ref{auto- of-b}), we get $AC_{b}(\frac{p-1}{2}-\tau^{\prime})=-AC_{b}(\tau^{\prime})$. Particularly, for $p\equiv1\ \pmod{4}$ and $\tau^{\prime}=\frac{p-1}{4}$, we get $AC_{b}(\frac{p-1}{4})=-AC_{b}(\frac{p-1}{4})$, which implies $AC_{b}(\frac{p-1}{4})=0$.\\
(3) Since the pair $(c,-c)$ always gives different LSBs for $c\in \mathbb{F}_{p}^{\ast}$, the result follows.\ \  \ \ \ \ \ \ \ \ \ \ \ \ \ \ \ \ \ \ \ \ \ \ \ \ \ \ \ \ \ \ $\blacksquare$

In convenience, we always use $AC_b(I)=(AC_b(i))_{i\in I}$ , where
\begin{eqnarray}
I = \left\{ \begin{array}{ll}
\{1,2,\cdots,\frac{p-5}{4}\}, & \mathrm{for}\ p\equiv1~(\mathrm{mod}\ {4});\\
\{1,2,\cdots,\frac{p-3}{4}\}, & \mathrm{for}\ p\equiv3~(\mathrm{mod}\ {4}).\\
\end{array}
\right.\label{DefI}
\end{eqnarray}


We note that $I=\emptyset$ for $p=3, 5$. Based on all the lemmas above, we obtain the following result.
\begin{table}
\centering
{\tiny
\begin{tabular}{|c|c|c|}
\hline$p$&$\beta$&$AC_b(I)$\\
\hline3&2&
$\emptyset$
\\
\hline5&2,3&
$\emptyset$
\\
\hline7&3,5&
$(2)$
\\
\hline11&2,6,7,8&
$(-2,2)$
\\
\hline13&2,6,7,11&
$(0,-4)$
\\
\hline17&3&
$(4,0,-4)$
\\
\hline19&2&
$(-2,2,-2,-6)$
\\
\hline23&5&
$(2,-2,2,-2,-6)$
\\
\hline29&2&
$(0,-4,0,-4,8,4)$
\\
\hline31&3&
$(10,6,2,-2,-6,-2,2)$
\\
\hline37&2&
$(0,-4,0,4,-8,4,0,-12)$
\\
\hline41&6&
$(4,-8,4,0,-12,0,4,0,4)$
\\
\hline43&3&
$(14,2,-2,-6,-2,2,6,2,-2,2)$
\\
\hline47&5&
$(10,-2,-14,-2,2,6,2,-2,-6,-2,2)$
\\
\hline53&2&
$(0,-4,0,-4,8,4,0,-4,-16,4,0,-4)$
\\
\hline59&2&
$(-2,2,-2,-6,-2,10,-2,18,-2,2,-10,2,-2,2)$
\\
\hline61&2&
$(0,-4,0,-4,0,20,0,-12,0,-4,-8,4,0,-4)$
\\
\hline67&2&
$(-2,2,-2,2,-2,-22,6,2,-2,-6,-2,10,-2,-6,14,2)$
\\
\hline71&7&
$(10,6,2,-10,2,-2,-14,-2,-22,-2,2,-2,2,-2,2,6,-6)$
\\
\hline73&5&
$(12,0,-12,0,4,24,4,0,-4,0,4,8,4,0,-4,0,4)$
\\
\hline79&3&
$(26,6,2,-2,-6,-2,-6,-2,2,6,2,-2,-6,-10,2,14,2,-2,2)$
\\
\hline83&2&
$(-2,2,-2,-6,6,-6,-2,10,-2,26,-2,2,-2,-14,-2,2,-2,2,-2,10)$
\\
\hline89&3&
$(28,8,4,8,4,8,12,0,-4,0,4,0,-4,0,-4,0,4,16,4,0,-4)$
\\
\hline 97&5&
$(20,0,-4,-8,4,0,4,0,-4,8,4,0,4,0,4,0,-12,0,4,0,-4,-32,-12)$
\\
\hline
\end{tabular}
\caption{\bf{Examples of $AC_b(I)$ for primes less than 100}} \label{table1}
}
\end{table}

\begin{theorem}\label{proper of auto}
Let the symbols be defined as before. Then, for $0<\tau<N$, the autocorrelation of the LSB sequence $\{s_t\}_{t=0}^{N-1}$ of a $p$-ary $m$-sequence $\{a_t\}_{t=0}^{N-1}$ is expressed as
\begin{equation}
AC_{s}(\tau)=\left\{
\begin{array}{lllll}
\big(1+AC_b(\tau^{\prime})\big)p^{n-1}-1,\ \mathrm{if}\ \tau\in\big\{M\tau^{\prime}\mid\tau^{\prime}\in I\big\}\cup\big\{M(p-1-\tau^{\prime})\mid\tau^{\prime}\in I\big\};\\
\big(1-AC_b(\tau^{\prime})\big)p^{n-1}-1,\ \mathrm{if}\ \tau\in\big\{M(\frac{p-1}{2}-\tau^{\prime})\mid \tau^{\prime}\in I\big\}\cup\big\{M(\frac{p-1}{2}+\tau^{\prime})\mid \tau^{\prime}\in I\big\};\\
p^{n-1}-1,\ \ \ \ \ \ \ \ \ \ \ \ \ \  \ \ \  \ \mathrm{if}\ p\equiv 1 ~(\bmod{~ 4})\ \mathrm{and}\ \tau=\frac{p^n-1}{4};\\
-(p-2)p^{n-1}-1,\ \ \ \ \ \ \ \ \mathrm{if}\ \tau=\frac{p^n-1}{2}; \\
p^{n-2}-1,\ \ \ \ \ \ \ \ \ \ \ \ \ \ \ \ \ \ \mathrm{otherwise}.
\end{array}
\right.\label{general auto}
\end{equation}
In particular, the corresponding autocorrelations $AC_{s}(\tau)$ for $p=3$ and $p=5$ can be given directly by
\begin{eqnarray}
AC_{s}(\tau)&=&\left\{
\begin{array}{ll}
-3^{n-1}-1, \ \ \ \ \ \ \mathrm{if}\ \tau=M,\\
3^{n-2}-1,\ \ \ \ \ \ \ \ \mathrm{otherwise},
\end{array}
\right.\label{ternary auto}\\
AC_{s}(\tau)&=&\left\{
\begin{array}{ll}
5^{n-1}-1, \ \ \ \ \ \ \ \ \mathrm{if}\ \tau=M \ \mathrm{or}\ 3M,\\
-3\times5^{n-1}-1, \ \mathrm{if}\ \tau=2M,\\
5^{n-2}-1,\ \ \ \ \ \ \ \ \mathrm{otherwise}
\end{array}
\right.\label{5-ary auto}
\end{eqnarray}
respectively.\ \ \ \ \ \ \ \ \ \ \ \ \ \ \ \ \ \ \ \ \ \ \ \ \ \ \ \ \ \ \ \ \ \ \ \ \ \ \ \ \ \ \ \ \ \ \ \ \ \ \ \ \ \ \ \ \ \ \ \ \ \ \ \ \ \ \ \ \ \ \ \ \ \ \ \ \ \ \ \ \ \ \ \ \ \ \ \ \ \ \ \ \ \ \ \ \ \ \ \ \ \ \ \ \ \ \ \ \ \ \ \ \ \ \ \ \ \ \ \ $\blacksquare$
\end{theorem}
\begin{remark}
For the autocorrelation function $AC_b(\tau^{\prime})$ of the Costas sequence $\{b_j\}_{j=0}^{p-2}$ of period $p-1$, we have reduced its values from a set $\{AC_{b}(\tau)\}_{\tau=1}^{p-2}$ to a set $\{AC_{b}(\tau)|\tau\in I\}$. Hence the size of the problem is simplified to a quarter of the original size and it can be determined relatively more efficiently by computer. Indeed, we present the corresponding ordered array $AC_{b}(I)$ for all odd primes smaller than 100 in Table \ref{table1}. Moreover, by plugging the values of $AC_{b}(I)$ in Table 1 for each prime $7\leq p<100$ into the corresponding formula in Theorem \ref{proper of auto}, we can get the exact autocorrelation distribution of the LSB sequence of the corresponding $p$-ary $m$-sequence. Additionally, it can be observed from these examples that all the autocorrelation values satisfy $-\frac{p-1}{3}\leq AC_b(\tau^{\prime})\leq\frac{p-1}{3}$ for $\tau^{\prime}\in\{1,2,\cdots,p-2\}$ but $\tau^{\prime}\neq\frac{p-1}{2}$. Finding out the complete and theoretical result of the autocorrelation distribution of the Costas sequence $\{b_j\}_{j=0}^{p-2}$ will be an interesting research problem, but due to our limited ability we can not resolve it in this paper. So we sincerely invite those readers who are interested in this problem to participate in it.
\end{remark}
\begin{remark}
Also, from Theorem \ref{proper of auto}, it seems that the autocorrelation values of the LSB sequences are high, comparing to the periods of these sequences, which is bad for the security of a key stream sequence. However, since the period of the bit-component sequence used in the ZUC algorithm-the core of the 3GPP LTE International Encryption Standard is huge (here $p=2^{31}-1$ and the period $N=p^{16}-1$) and only a little part of the sequence is chosen to be as a key stream in the encryption process, then the high autocorrelation of the sequence has almost no negative impact on the security of the whole cipher system.
\end{remark}

\begin{theorem}\label{shift equi}
Let $p=2^k-1$ be a  Mersenne prime, and $\{a_{t,i-1}\}_{t=0}^{N-1}$  the $i$-th bit-component sequence of $\{a_t\}_{t=0}^{N-1}$. Then, for $2\leq i\leq k$, the $i$-th bit-component sequence $\{a_{t,i-1}\}_{t=0}^{N-1}$ is a cyclic shift of the LSB sequence $\{s_{t}\}_{t=0}^{N-1}$.
\end{theorem}
$\mathbf{Proof.}$ Because $2\in \mathbb{F}_{p}$, there exists some $1\leq j_0\leq p-2$ and $\tau_0=\frac{p^n-1}{p-1}j_0$ such that $2=\alpha^{\tau_0}$.
Then
$$
2a_{t}=2\mathrm{Tr}(\alpha^t)=\mathrm{Tr}(\alpha^{t+\tau_0})=a_{t+\tau_0},
$$
which shows that $\{2a_t\}$ is the left cyclic shift of $\{a_t\}$ by $\tau_0$ .
Moreover,
$$
2a_t\ \mathrm{mod}\ p=a_{t,k-1}+a_{t,0}\times2+a_{t,1}\times2^2+\cdots+a_{t,k-3}\times2^{k-2}+a_{t,k-2}\times2^{k-1},
$$
that is, the binary bit string of $2a_t$ is the left cyclic shift of the binary bit string of $a_t$ by 1. Therefore,  for $1\leq i\leq k$, the $((i\ \mathrm{mod}\ k$)+1)-th bit-component sequence is the left cyclic shift of the $i$-th bit-component sequence by $\tau_0$, which results in the conclusion. \ \ \ \ \ \ \ \ \ \ \ \ \ \ \ \ \ \ \ \ \ \ \ \ \ \ \ \ \ \ \ \ \ \ \ \ \ \ \ \ \ \ \ \ \ \ \ \ \ \ \ \ \ \ \ \ \ \ \ \ \ \ \ \ \ \ \ \ \ \ \ \ \ \ \ \ \ \ \ \ \ \ \ \ \ \ \ \ \ \ \ \ \ \ \ \ \ \ \ \ \ \ \ \ \ $\blacksquare$

\section{Lower bound on the 2-adic complexity of each of these LSB sequences for $p<20$} \label{section 4}

First we describe the method of Hu \cite{Hu Honggang} as the following lemma. 

\begin{lemma}\label{method}\cite{Hu Honggang}
Let $T(x)=\sum\limits_{t=0}^{N-1}(-1)^{s_t}x^t\in \mathbb{Z}[x]$. Then
\begin{eqnarray}
-2S(x)T(x^{-1})\equiv N+\sum\limits_{\tau=1}^{N-1}AC_{s}(\tau)x^{\tau}-T(x^{-1})\left(\sum\limits_{t=0}^{N-1}x^{t}\right)\bmod\Big(x^N-1\Big). \label{finatrans}
\end{eqnarray}
\end{lemma}

\begin{lemma}\label{general 2-adic}
Suppose that $n\geq2$ is a positive integer and $I$ is defined as in Eq. (\ref{DefI}).
Then we have
\begin{eqnarray}
&&S(2)T(2^{-1})\equiv-\frac{2^{\frac{N}{2}}-1}{2^{M}-1}\Big(p-1\Big)p^{n-2}\ \mathrm{mod}\ \Big(2^{\frac{N}{2}}-1\Big).\label{minus}\\
&&S(2)T(2^{-1})\equiv \left(\sum\limits_{\tau^{\prime}\in I}AC_b(\tau^{\prime})\Big(2^{M(\frac{p-1}{2}-\tau^{\prime})}-2^{M\tau^{\prime}}\Big)-(p-1)\right)p^{n-1}\ \mathrm{mod}\ \Big(2^{\frac{N}{2}}+1\Big),\label{add}
\end{eqnarray}
\end{lemma}
$\mathbf{Proof.}$ We only present the proof for the case of $p\equiv3 \pmod{4} $ and the other case is similar. Substituting Eq. (\ref{general auto}) in Theorem \ref{proper of auto} into Eq. (\ref{finatrans}) in Lemma \ref{method}, we have
\begin{eqnarray}
-2S(x)T(x^{-1})&\equiv& N+\sum\limits_{\tau\neq M\tau^{\prime},\tau^{\prime}=1,2,\cdots,p-2}\Big(p^{n-2}-1\Big)x^{\tau}+\sum\limits_{\tau^{\prime}=1}^{\frac{p-3}{4}}\bigg[\Big(1+AC_b(\tau^{\prime})\Big)p^{n-1}-1\bigg]x^{M\tau^{\prime}}\nonumber\\
&&+\sum\limits_{\tau^{\prime}=1}^{\frac{p-3}{4}}\bigg[\Big(1-AC_b(\tau^{\prime})\Big)p^{n-1}-1\bigg]x^{M(\frac{p-1}{2}-\tau^{\prime})}+\bigg[-(p-2)p^{n-1}-1\bigg]x^{\frac{N}{2}}\nonumber\\
&& +\sum\limits_{\tau^{\prime}=1}^{\frac{p-3}{4}}\bigg[\Big(1-AC_b(\tau^{\prime})\Big)p^{n-1}-1\bigg]x^{M(\frac{p-1}{2}+\tau^{\prime})}\nonumber\\
&&+\sum\limits_{\tau^{\prime}=1}^{\frac{p-3}{4}}\bigg[\Big(1+AC_b(\tau^{\prime})\Big)p^{n-1}-1\bigg]x^{M(p-1-\tau^{\prime})}-T\Big(x^{-1}\Big)\left(\sum\limits_{t=0}^{N-1}x^{t}\right)\bmod\Big(x^N-1\Big)\nonumber
\end{eqnarray}
\begin{eqnarray}
&\equiv &N-\Big(p^{n-2}-1\Big)+\sum\limits_{\tau=0}^{N-1}\Big(p^{n-2}-1\Big)x^{\tau}+\sum\limits_{\tau^{\prime}=1}^{\frac{p-3}{4}}\bigg[\Big(1+AC_b(\tau^{\prime})\Big)p^{n-1}-p^{n-2}\bigg]x^{M\tau^{\prime}}\\
&&+\sum\limits_{\tau^{\prime}=1}^{\frac{p-3}{4}}\bigg[\Big(1-AC_b(\tau^{\prime})\Big)p^{n-1}-p^{n-2}\bigg]x^{M(\frac{p-1}{2}-\tau^{\prime})}+\bigg[-\Big(p-2\Big)p^{n-1}-p^{n-2}\bigg]x^{\frac{N}{2}}\nonumber\\
&&+\sum\limits_{\tau^{\prime}=1}^{\frac{p-3}{4}}\bigg[\Big(1-AC_b(\tau^{\prime})\Big)p^{n-1}-p^{n-2}\bigg]x^{M(\frac{p-1}{2}+\tau^{\prime})}\nonumber\\
&&+\sum\limits_{\tau^{\prime}=1}^{\frac{p-3}{4}}\bigg[\Big(1+AC_b(\tau^{\prime})\Big)p^{n-1}-p^{n-2}\bigg]x^{M(p-1-\tau^{\prime})}-T\Big(x^{-1}\Big)\left(\sum\limits_{t=0}^{N-1}x^{t}\right)\bmod\Big(x^N-1\Big)\nonumber\\
&\equiv &\left\{\sum\limits_{\tau^{\prime}=1}^{\frac{p-3}{4}}\Bigg[\bigg(\Big(1+AC_b(\tau^{\prime})\Big)p-1\bigg)x^{M\tau^{\prime}}+\bigg(\Big(1-AC_b(\tau^{\prime})\Big)p-1\bigg)x^{M(\frac{p-1}{2}+\tau^{\prime})}\Bigg]\right.\nonumber\\
&&\left.+\sum\limits_{\tau^{\prime}=1}^{\frac{p-3}{4}}\Bigg[\bigg(\Big(1-AC_b(\tau^{\prime})\Big)p-1\bigg)x^{M(\frac{p-1}{2}-\tau^{\prime})}+\bigg(\Big(1+AC_b(\tau^{\prime})\Big)p-1\bigg)x^{M(p-1-\tau^{\prime})}\Bigg]\right\}p^{n-2}\nonumber\\
&&+\Big(p^{2}-1\Big)p^{n-2}-\Big(p-1\Big)^2p^{n-2}x^{\frac{N}{2}}-\Big(p^{n-2}-1+T(x^{-1})\Big)\left(\sum\limits_{t=0}^{N-1}x^{t}\right)\bmod\Big(x^N-1\Big).\nonumber
\end{eqnarray}
Furthermore, we note that $x^{M\times\frac{p-1}{2}}=x^{\frac{N}{2}}\equiv1\bmod(x^{\frac{N}{2}}-1)$ and $x^{\frac{N}{2}}\equiv-1\bmod(x^{\frac{N}{2}}+1)$. Substituting $x$ for 2, the desirable results can be derived.
\ \ \ \ \ \ \ \ \ \ \ \ \ \ \ \ \ \ \ \ \ \ \ \ \ \ \ \ \ \ \ \ \ \ \ \ \ \ \ \ \ \ \ \ \ \ \ \ \ \ \ \ \ \ \ \ \ \ \ \ \ \ \ \ \ \ \ \ \ \ \ \ \ \ \ \ $\blacksquare$

In the sequel, we also need the following result from the elementary number theory.
\begin{lemma}\label{genel-prime-prop}
(1) Let $p$ be an odd prime and $n$ a positive integer. Then $p \mid (2^{p^n-1}-1)$. Furthermore, $p^2\mid (2^{p^n-1}-1)$ if and only if $p$ is a Wieferich prime (An odd prime $p$ satisfying $p^{2}\mid (2^{p-1}-1)$ is called a Wieferich prime. It is shown in \cite{Dorais} that there are only two Wieferich primes 1093 and 3511 up to $6.7\times 10^{15}$).\\
(2) A Mersenne prime $p=2^k-1$ is not a Wieferich prime. Furthermore, for an odd prime $k$, we have $p \mid (2^{\frac{p^n-1}{2}}-1)$, $p^{2}\nmid(2^{\frac{p^n-1}{2}}-1)$, $p\nmid(2^{\frac{p^n-1}{2}}+1)$.
\end{lemma}
$\mathbf{Proof.}$
(1) Due to $(p-1) \mid (p^n-1)$, we have $(2^{p-1}-1) \mid (2^{p^n-1}-1)$. By Fermat's little Theorem we know that $p \mid (2^{p-1}-1)\Rightarrow p \mid (2^{p^n-1}-1)$. By Euler's theorem, we have $2^{\phi(p^2)}=2^{p(p-1)}\equiv1 ~(\mathrm{mod}\ p^2)$, where $\phi(\cdot)$ is Euler's phi Function. And $p^n-1=(p-1)(p^{n-1}+p^{n-2}+\cdots+p+1)\equiv p-1\ (\mathrm{mod}\ (p(p-1)))$, which implies that $2^{p^n-1}-1\equiv2^{p-1}-1\ (\mathrm{mod}\ p^2)$. Therefore, $p^2\mid (2^{p^n-1}-1)$ if and only if $p^2\mid (2^{p-1}-1)$.\\
(2) Notice that $k=2$ or $k$ is an odd prime for a Mersenne prime $p=2^k-1$. If $k=2$, i.e., $p=3$, then $p$ is not a Wieferich prime from the conclusion in \cite{Dorais}.
If $k$ is an odd prime for $p=2^k-1$, we get $k\mid (p-1)$ by $p\mid (2^{p-1}-1)$. Suppose $p^2 \mid (2^{p-1}-1)$, i.e., $\left(2^k-1\right)^2 \mid \Big[\left(2^k-1\right)\left(2^{(\frac{p-1}{k}-1)k}+2^{(\frac{p-1}{k}-2)k}+\cdots+2^k+1\right)\Big]$,
which implies
\begin{equation}
(2^k-1) \mid( 2^{(\frac{p-1}{k}-1)k}+2^{(\frac{p-1}{k}-2)k}+\cdots+2^k+1).\label{contra}
\end{equation}
 But we know $$2^{(\frac{p-1}{k}-1)k}+2^{(\frac{p-1}{k}-2)k}+\cdots+2^k+1\equiv\frac{p-1}{k}\equiv\frac{2(2^{k-1}-1)}{k}\ \mathrm{mod}\ (2^k-1)$$
 and
 $\mathrm{gcd}\left(2(2^{k-1}-1),2^{k}-1\right)=1$, i.e., $\mathrm{gcd}(2^k-1,2^{(\frac{p-1}{k}-1)k}+2^{(\frac{p-1}{k}-2)k}+\cdots+2^k+1)=1$,
 a contradiction to the Eq. (\ref{contra}). Hence $p$ is not a Wieferich prime. Furthermore, since $k \mid (2^{k-1}-1)$, $2^{k-1}-1=\frac{p-1}{2}$ and $\frac{p-1}{2} \mid \frac{p^n-1}{2}$, we get $k \mid \frac{p^n-1}{2}$ and $(2^k-1) \mid (2^{\frac{p^n-1}{2}}-1)$, i.e., $p \mid (2^{\frac{p^n-1}{2}}-1)$. Moreover, $p^{2}\nmid(2^{N}-1)$ implies that $p^2\nmid(2^{\frac{N}{2}}-1)$ and $p \mid (2^{\frac{p^n-1}{2}}-1)$ results in $p\nmid(2^{\frac{p^n-1}{2}}+1)$. \ \ \ \ \ \ \ \ \ \ \ \ \ \ \ \ \ \ \ \ \ \ \ \ \ \ \ \ \ \ \ \ \ \ \ \ \ \ \ \ $\blacksquare$
\begin{lemma}\label{general 2-adic}
Let the notations be the same as above and let $\delta:=\mathrm{Ord}_{p}(2)$ be the multiplicative order of 2 modular $p$. Suppose that $n\geq2$ is a positive integer. Then we have the following two results:\\
(1)
{\small
\begin{eqnarray}
\mathrm{gcd}\left(S(2)T(2^{-1}),2^{\frac{N}{2}}-1\right)&=&\left\{
\begin{array}{ll}
\mathrm{gcd}\Big((p-1)p^{n-2},2^M-1\Big)\frac{2^{\frac{N}{2}}-1}{2^M-1}, &  \mathrm{if}~n\equiv0\ (\mathrm{mod}\ \delta ),  n\neq2, \\
\mathrm{gcd}\Big(p-1,2^M-1\Big)\frac{2^{\frac{N}{2}}-1}{2^M-1}, & \mathrm{if} ~n\not\equiv0\ (\mathrm{mod}\ \delta)\ \mathrm{or}\ n=2,
\end{array}
\right.\label{nonMersenne-prime-2}\\
\mathrm{gcd}\left(S(2)T(2^{-1}),2^{\frac{N}{2}}+1\right)&=&\left\{
\begin{array}{llll}
\mathrm{gcd}\left(p^{n-1}\bigg(\sum\limits_{\tau^{\prime}\in I}AC_b(\tau^{\prime})\Big(2^{M(\frac{p-1}{2}-\tau^{\prime})}-2^{M\tau^{\prime}}\Big)-(p-1)\bigg),2^{\frac{N}{2}}+1\right),\\
\ \ \ \ \ \  \ \ \ \  \ \ \ \ \ \ \ \ \ \ \ \ \  \ \ \ \ \ \ \ \ \ \ \ \ \ \ \ \ \ \ \ \ \mathrm{if}\ \delta \nmid\frac{p-1}{2} \mathrm{and}\ n\ is\ \mathrm{odd},\\
\mathrm{gcd}\left(\sum\limits_{\tau^{\prime}\in I}AC_b(\tau^{\prime})\Big(2^{M(\frac{p-1}{2}-\tau^{\prime})}-2^{M\tau^{\prime}}\Big)-(p-1),2^{\frac{N}{2}}+1\right),\\
\ \ \ \ \ \ \ \ \ \ \ \ \ \ \ \ \  \ \ \ \ \ \ \ \ \ \ \ \ \ \ \ \ \ \ \ \ \ \ \ \ \ \ \ \mathrm{if}\ \delta \mid \frac{p-1}{2}\ \mathrm{or}\ n\ \mathrm{is\ even}.
\end{array}
\right.\label{nonMersenne-prime-1}
\end{eqnarray}
}
Particularly, for $p=3,5$, we have
\begin{eqnarray}
\mathrm{gcd}\left(S(2)T(2^{-1}),\ 2^N-1\right)&=&\left\{
\begin{array}{ll}
1,\ \ \ \ \ \ \ \ \ \ \ \mathrm{if}\ n=2,\\
3,\ \ \ \ \ \ \ \ \ \ \ \mathrm{if}\ n>2,
\end{array}
\right.\label{special-prime-1}\\
\mathrm{gcd}\left(S(2)T(2^{-1}),\ 2^N-1\right)&=&\left\{
\begin{array}{ll}
2^M+1,\ \ \ \ \mathrm{ if}\ n\equiv2\pmod{4},\\
5(2^M+1),\ \mathrm{otherwise}
\end{array}
\right.\label{special-prime-2}
\end{eqnarray}
respectively.\\
(2) If $p=2^k-1>5$ is a Mersenne prime, then
{\small
\begin{eqnarray}
\mathrm{gcd}\left(S(2)T(2^{-1}),2^{\frac{N}{2}}-1\right)=\left\{
\begin{array}{ll}
\mathrm{gcd}\Big(p-1,2^M-1\Big)\frac{2^{\frac{N}{2}}-1}{2^M-1}p, \ \ \mathrm{if}~ n\equiv0\ (\mathrm{mod}\ \delta)\ \mathrm{but}\ n\neq2,\\
\mathrm{gcd}\Big(p-1,2^M-1\Big)\frac{2^{\frac{N}{2}}-1}{2^M-1}, \ \ \ \mathrm{if}~  n\not\equiv0\ (\mathrm{mod}\ \delta)\ \mathrm{or}\ n=2,
\end{array}
\right.\label{Mersenne-prime-2}
\end{eqnarray}
}
{\small
\begin{eqnarray}
\mathrm{gcd}\left(S(2)T(2^{-1}),2^{\frac{N}{2}}+1\right)&=&\left\{
\begin{array}{llll}
\mathrm{gcd}\left(p\bigg(\sum\limits_{\tau^{\prime}\in I}AC_b(\tau^{\prime})\Big(2^{M(\frac{p-1}{2}-\tau^{\prime})}-2^{M\tau^{\prime}}\Big)-(p-1)\bigg),2^{\frac{N}{2}}+1\right),\\
\ \ \ \ \ \ \ \  \ \ \ \ \ \ \  \ \ \  \ \ \ \ \ \ \ \ \ \ \ \ \ \ \ \ \ \mathrm{if}\ \delta\nmid\frac{p-1}{2}\ \mathrm{and}\  n\ \mathrm{is\ odd}, \\
\mathrm{gcd}\left(\sum\limits_{\tau^{\prime}\in I}AC_b(\tau^{\prime})\Big(2^{M(\frac{p-1}{2}-\tau^{\prime})}-2^{M\tau^{\prime}}\Big)-(p-1),2^{\frac{N}{2}}+1\right),\\
\ \ \ \ \ \ \ \ \ \ \ \ \ \ \ \ \ \ \ \ \ \ \ \ \  \  \ \ \  \ \ \ \ \ \ \mathrm{if}\ \delta\mid \frac{p-1}{2}\ \mathrm{or}\ n\ \mathrm{is\ even.}
\end{array}
\right.\label{Mersenne-prime-1}
\end{eqnarray}
}
\end{lemma}
$\mathbf{Proof.}$ (1) From Eq. (\ref{minus}), we get
{\small
$\mathrm{gcd}\left(S(2)T(2^{-1}),2^{\frac{N}{2}}-1\right)=\frac{2^{\frac{N}{2}}-1}{2^{M}-1}\mathrm{gcd}\Big((p-1)p^{n-2},\ 2^{M}-1\Big)$.
}
Note that $2^{p^i}\equiv2~(\bmod~ p)$ for any nonnegative integer $i$ by Fermat's Little Theorem. Since $M=\frac{N}{p-1}=p^{n-1}+p^{n-2}+\cdots+p+1$,
we get $2^M=2^{p^{n-1}+p^{n-2}+\cdots+p+1}\equiv2^{n}~(\bmod ~p)$. By the definition of $\delta$, we know that $2^M-1\equiv0\ (\mathrm{mod}\ p)$ if $n\equiv0\ (\mathrm{mod}\ \delta)$, otherwise, $2^M-1\not\equiv0\ (\mathrm{mod}\ p)$, the Eq. (\ref{nonMersenne-prime-2}) holds. \\
Similarly, since $2^{\frac{N}{2}}=(2^{M})^{\frac{p-1}{2}}\equiv2^{\frac{n(p-1)}{2}}~(\bmod ~p)$ and $2^{p-1}\equiv1\ (\mathrm{mod}\ p)$ by Fermat Little Theorem,
we can get $2^{\frac{p-1}{2}}\equiv-1~(\bmod ~p)$ if $\delta\nmid\frac{p-1}{2}$. Combining Eq. (\ref{add}), the Eq. (\ref{nonMersenne-prime-1}) holds.

Particularly, for $p=3$, we have $N=2M$ and $I=\emptyset$ by Theorem \ref{proper of auto}.
Then, $\mathrm{gcd}(p-1,\ 2^M-1)=\mathrm{gcd}(2,\ 2^M-1)=1$, $\mathrm{gcd}(p-1,\ 2^{\frac{N}{2}}+1)=\mathrm{gcd}(2,\ 2^{\frac{N}{2}}+1)=1$, $\mathrm{Ord}_3(2)=2$, and $\mathrm{Ord}_3(2)\nmid\frac{3-1}{2}$.
Since 3 is not a Wieferich prime, we get $3^2\nmid(2^{M}-1)$ and $3^2\nmid(2^{\frac{N}{2}}+1)$ from Lemma \ref{genel-prime-prop}, i.e., $\mathrm{gcd}(3^{n-2},\ 2^{M}-1)=3$ for an even $n>2$, and $\mathrm{gcd}(3^{n-1},\ 2^{\frac{N}{2}}+1)=3$ for odd $n>1$. Hence, combining Eqs. (\ref{nonMersenne-prime-2}-\ref{nonMersenne-prime-1}), the Eq. (\ref{special-prime-1}) can be proved. Similarly, Eq. (\ref{special-prime-2}) for $p=5$ can also be derived.\\
(2) The proof is similar to the above.\ \ \ \ \ \ \ \ \ \ \ \ \ \ \ \ \ \ \ \ \ \ \ \ \ \ \ \ \ \ \ \ \ \ \ \ \ \ \ \ \ \ \ \ \ \ \ \ \ \ \ \ \ \ \ \ \ \ \ \ \ \ \ \ \ \ \ \ \ \ \ \ \ \ \ \ \ \ \ \ \ \ \ \ \ \ \ \ \ \ \ \ \ \ \ \ \ \ \ \ \ \ \ \ \ \ \ \ \ \ \ \ \ \ \ \ \ \ \ $\blacksquare$
\begin{theorem}\label{ternary 2-adic}
Let $\{s_t\}_{t=0}^{N-1}$ be the LSB sequence of a ternary $m$-sequence of order $n\geq2$. Then the 2-adic complexity $\Phi_{2}(s)$ is bounded by
$\Phi_{2}(s)\geq N-3$.
\end{theorem}
$\mathbf{Proof.}$ From the Eqs. (\ref{2-adic calculation}) and (\ref{special-prime-1}), the 2-adic complexity of $\{s_t\}_{t=0}^{3^n-2}$ satisfies
\begin{eqnarray}
\Phi_{2}(s)&=\lfloor \mathrm{log}_2\frac{2^N-1}{\mathrm{gcd}\left(S(2), 2^N-1\right)}\rfloor\geq\lfloor \mathrm{log}_2\frac{2^N-1}{\mathrm{gcd}\left(S(2)T(2^{-1}), 2^N-1\right)}\rfloor\nonumber\\
&\geq N-1-\lceil \mathrm{log}_2\mathrm{gcd}\left(S(2)T(2^{-1}), 2^N-1\right)\rceil\geq N-3.\nonumber\ \ \ \ \ \ \ \ \ \ \ \ \ \ \ \ \ \ \ \ \ \ \ \ \ \ \ \ \ \ \ \ \ \ \ \ \ \ \ \ \ \ \ \ \ \ \ \ \ \ \ \ \ \ \blacksquare
\end{eqnarray}

\begin{theorem}\label{5-ary 2-adic}
Let $\{s_t\}_{t=0}^{N-1}$ be the LSB sequence of a 5-ary $m$-sequence of order $n\geq2$. Then the 2-adic complexity $\Phi_{2}(s)$ is bounded by $\Phi_{2}(s)\geq \frac{3N}{4}-5$.
\end{theorem}
$\mathbf{Proof.}$ The proof is similar to that of Theorem \ref{ternary 2-adic} except that we  use Eq. (\ref{special-prime-2}) in Lemma \ref{general 2-adic} instead. \ \ $\blacksquare$

In fact, we can also derive a lower bound on the 2-adic complexity of the LSB sequence for $p=7,11,13,17,19$ respectively. The proofs are similar to the cases of $p=3,5$ except that we need to use Euclid Algorithm when determining the corresponding values of Eqs.(\ref{nonMersenne-prime-1}) and (\ref{Mersenne-prime-1}). In order to avoid repetition, we skip and present them in Table~\ref{table-2} for these results.

\begin{table}
\centering
\begin{tabular}{|c|c|c|}
\hline$p$&$\beta$&the lower bound on the 2-adic complexity $\Phi_2(s)$\\
\hline3&2&
$\Phi_2(s)\geq N-3=\frac{N}{2}+\frac{N}{2}-3=\frac{N}{2}+\frac{N}{p-1}-3$
\\
\hline5&2,3&
$\Phi_2(s)\geq \frac{3N}{4}-5=\frac{N}{2}+\frac{N}{4}-5=\frac{N}{2}+\frac{N}{p-1}-5$
\\
\hline7&3,5&
$\Phi_2(s)\geq \frac{2N}{3}-7=\frac{N}{2}+\frac{N}{6}-7=\frac{N}{2}+\frac{N}{p-1}-7$
\\
\hline11&2,6,7,8&
$\Phi_2(s)\geq \frac{3N}{5}-8=\frac{N}{2}+\frac{N}{10}-8=\frac{N}{2}+\frac{N}{p-1}-8$
\\
\hline13&2,6,7,11&
$\Phi_2(s)\geq \frac{7N}{12}-4=\frac{N}{2}+\frac{N}{12}-4=\frac{N}{2}+\frac{N}{p-1}-4$
\\
\hline17&3&
$\Phi_2(s)\geq \frac{9N}{16}-7=\frac{N}{2}+\frac{N}{16}-7=\frac{N}{2}+\frac{N}{p-1}-7$
\\
\hline19&2&
$\Phi_2(s)\geq \frac{5N}{9}-15=\frac{N}{2}+\frac{N}{18}-15=\frac{N}{2}+\frac{N}{p-1}-15$
\\
\hline
\end{tabular}
\caption{\bf{Examples of $\Phi_2(s)$ for $p\leq 19$}} \label{table-2}
\end{table}
\begin{theorem}\label{7ary-2-adic}
Let $p=7$, $n\geq2$ a positive integer, and $\{s_t\}_{t=0}^{N-1}$ the LSB sequence of any 7-ary $m$-sequence of order $n$. Then the 2-adic complexity $\Phi_{2}(s)$ of $\{s_t\}_{t=0}^{N-1}$ satisfies $\Phi_{2}(s)\geq\frac{2N}{3}-7$. \ \ \ \ \  \ \ \ \ \ \ \ \ \ \ \ \ \ \ \ \ \ $\blacksquare$
\end{theorem}
\begin{theorem}\label{11-ary 2-adic}
Let $p=11$, $n\geq2$ a positive integer, and $\{s_t\}_{t=0}^{N-1}$ the LSB sequence of any 11-ary $m$-sequence of order $n$. Then the lower bound on the 2-adic complexity $\Phi_{2}(s)$ of $\{s_t\}_{t=0}^{N-1}$ is given by $\Phi_{2}(s)\geq \frac{3N}{5}-8$.\ \ \ \ \  \ \ \ \ \ \ \ \ \ \ \ \ \ \ \ \ \ \ \ \ \ \  \ \ \ \ \ \ \ \ \ \ \ \ \ \ \ \ \ \ \ \ \ \ \ \ \ \ \ \ \ \ \ \ \ \ \ \ \ \ \ \ \ \ \ \ \ \ \ \ \ \ \ \ \ \ \ \ \ \ \ \ \ \ \ \ \ \ \ \ \ \ \ \ \ \ \ \ \ \ \ \ \ \ \ \ \ $\blacksquare$
\end{theorem}
\begin{theorem}\label{13-ary 2-adic}
Let $p=13$, $n\geq2$ a positive integer, and $\alpha$ be a primitive element of $\mathbb{F}_{13^n}$ such that $\beta=\alpha^M=2,6,7$ or $11$.
Let  $\{s_t\}_{t=0}^{N-1}$ be the LSB sequence of the 13-ary $m$-sequence defined by $\alpha$. Then the 2-adic complexity $\Phi_{2}(s)$ of $\{s_t\}_{t=0}^{N-1}$ satisfies $\Phi_{2}(s)\geq \frac{7N}{12}-4$.\ \ \ \ \  \ \ \ \ \ \ \ \ \ \ \ \ \ \ \ \ \ \ \ \ \ \  \ \ \ \ \ \ \ \ \ \ \ \ \ \ \ \ \ \ \ \ \ \ \ \ \ \ \ \ $\blacksquare$
\end{theorem}
\begin{theorem}\label{17-ary 2-adic}
Let $p=17$, $n\geq2$ a positive integer, and $\alpha$ be a primitive element of $\mathbb{F}_{17^n}$ such that $\beta=\alpha^M=3$.
Let  $\{s_t\}_{t=0}^{N-1}$ be the LSB sequence of the 17-ary $m$-sequence defined by $\alpha$. Then the 2-adic complexity $\Phi_{2}(s)$ of $\{s_t\}_{t=0}^{N-1}$ satisfies $\Phi_{2}(s)\geq \frac{9N}{16}-7$.\ \ \ \ \  \ \ \ \ \ \ \ \ \ \ \ \ \ \ \ \ \ \ \ \ \ \ \ \ \ \ \ \ \ \ \  \ \ \ \ \ \ \ \ \ \ \ \ \ \ \ \ \ \ \ \ \ $\blacksquare$
\end{theorem}
\begin{theorem}\label{19-ary-2-adic}
Let $p=19$, $n\geq2$ a positive integer, and $\alpha$ be a primitive element of $\mathbb{F}_{19^n}$ such that $\beta=\alpha^M=2$. Let $\{s_t\}_{t=0}^{N-1}$ be the LSB sequence of the 19-ary $m$-sequence defined by $\alpha$. Then the lower bound on the 2-adic complexity $\Phi_{2}(s)$ of $\{s_t\}_{t=0}^{N-1}$ is given by $\Phi_{2}(s)\geq\frac{5N}{9}-15$.\ \  \ \ \ \ \ \ \ \ \ \ \ \ \ \ \ \ \ \ \ \ \ \ \ \ \ \ \ \ $\blacksquare$
\end{theorem}
\begin{remark}
In the process of computing the lower bound on the 2-adic complexity of each LSB sequence of the above six classes, we always suppose $n\geq2$. In fact, it can be testified by simply calculation that all the lower bounds also hold for $n=1$.
\end{remark}
\begin{remark}
In order to observe the laws of the 2-adic complexity of the LSB sequence in each of Theorems \ref{ternary 2-adic}-\ref{19-ary-2-adic}, we list the Table \ref{table-2}, from which it is obvious that, for $n\geq2$, the main part in the expression of the lower bound of the 2-adic complexity of the LSB sequence (all the bit-component sequences for a Mersenne prime) of the $p$-ary $m$-sequence for each $p\leq 19$ have a unified form, i.e., $\frac{N}{2}+\frac{N}{p-1}$, which are large enough to resist the RAA. Not only so, and we can also get a similar lower bound on the 2-adic complexity of the LSB sequence of the $p$-ary $m$-sequence through similar method for $p$ taking 17, 23, 29, 31 repectively. Therefore, we give the following conjecture.
\end{remark}
By Table~\ref{table-2}, we propose the following conjecture.
\begin{conjecture}
Let $p$ be any odd prime, $n$ a positive integer, $N=p^n-1$, and $\{s_t\}_{t=0}^{N-1}$ the LSB sequence of a $p$-ary $m$-sequence of order $n$. Then the 2-adic complexity $\Phi_{2}(s)$ of $\{s_t\}_{t=0}^{N-1}$ is lower bounded by
$\frac{p+1}{2(p-1)}N-C_{p}$ which is larger than $\frac{N}{2}$ when $n\geq2$, where the constant number $C_{p}$ depends only on $p$.
\end{conjecture}

\section*{Data Availability}
No data were used to support this study.
\section*{Conflicts of Interest}
The authors declare that they have no conflicts of interest.

\section*{Acknowledgement}

Yuhua Sun is financially supported by the National Natural Science Foundation of China (No. 61902429, No.11775306), Fundamental Research Funds for the Central Universities (No. 19CX02058A), Shandong Provincial Natural Science Foundation of China (No. ZR2017MA001, ZR2019MF070), the Open Research Fund from Shandong provincial Key Laboratory of Computer Networks, Grant No. SDKLCN-2018-02, Key Laboratory of Applied Mathematics of Fujian Province University (Putian University)(No. SX201806).

Qiuyan Wang is supported by the National Natural Science Foundation of China (No. 61602342), the Natural Science Foundation of Tianjin (No. 18JCQNJC70300), the Science and Technology Development Fund of Tianjin Education Commission for Higher Education (No. 2018KJ215), the Key Laboratory of Applied Mathematics of Fujian Province University (Putian University) (No. SX201804, No. SX201904), the China Scholarship Council (No. 201809345010), NSFC (No. 61972456, No. 61802281, No. 2017KJ237).


\begin{thebibliography}{1}


\bibitem{Agnes Hui} A. H. Chan and  R. A. Games, \lq\lq On the Linear Span of Binary Sequences Obtained from Finite Geometries.
\rq\rq {\em Advances in Cryptology-CRYPTO'86,} LNCS 263, pp. 405-417, 1987.

\bibitem{Ding Cunsheng} C. Ding, T. Helleseth, W. Shan, \lq\lq On the linear complexity of Legendre sequences.\rq\rq {\em IEEE Trans. Inf. Theory,} vol. 45, no. 2, pp. 693-698, 1998.

\bibitem{Dorais} F. Dorais, D. Klyve, \lq\lq A Wieferich prime search up to $6.7\times 10^{15}$.\rq\rq {\em J. Integer Seq.,} vol. 14, no. 9, pp. 1-14, 2011.

\bibitem{Konstantinos} K. Drakakis, V. Requena, G. McGuire, \lq\lq On the nonlinearity of exponential Welch Costas Functions.\rq\rq {\em IEEE Trans. Inf. Theory,} vol. 56, no. 3, pp. 1230-1238, 2010.


\bibitem{Edemskiy} V. Edemskiy, A. Palvinskiy, \lq\lq The linear complexity of binary sequences of length $2p$ with optimal three-level autocorrelation.\rq\rq {\em Inform. Process. Lett.,} vol. 116, no. 2, pp. 153-156, 2016.

\bibitem{3GPP-1}ETSL/SAGE Specification. Specification of the 3GPP confidentiality and integrity algorithms 128-EEA3 \& 128-ELA3. Document 2: ZUC Specification, Version: 1.6, (2011).

\bibitem{3GPP-2} ETSL/SAGE Specification, Specification of the 3GPP confidentiality and integrity algorithms 128-EEA3 \& 128-EIA3. Document 1: 128-EEA3 and 128-EIA3 Specification; Version: 1.6, (2011).

\bibitem{Etzion} T. Etzion, \lq\lq Linear complexity of de Bruijn sequences-old and new results.\rq\rq {\em IEEE Trans. Inf. Theory,} vol. 45, no. 2, pp. 693-698, 1999.

\bibitem{Golomb} S. W. Golomb, \lq\lq  Algebraic constructions for Costas arrays.\rq\rq  {\em J. Combin. Theory Ser. A  } 37 (1984), no. 1, 13-21.

\bibitem{Helleseth} T. Helleseth, J. E. Mathiassen, M. Maas, T. Segers, \lq\lq Linear complexity over $\mathbb{F}_p$ of Sidel'nikov sequences.\rq\rq {\em ISIT 2004,} pp. 122, 2004.

\bibitem{Hofer} R. Hofer and A. Winterhof, \lq\lq On the 2-adic complexity of the two-prime generator.\rq \rq {\em IEEE Trans. Inf. Theory,} to appear.

\bibitem{Hu Honggang} H. Hu, \lq\lq Comments on \lq A New Method to Compute the 2-Adic Complexity of Binary Sequences\rq. \rq\rq {\em IEEE Trans. Inf. Theory,} vol. 60, no. 9, pp. 5803-5804, 2014.

\bibitem{Hu Liqin} L. Hu, Q. Yue, M. Wang, \lq\lq The Linear Complexity of Whiteman's Generalized Cyclotomic Sequences of Period $p^{m+1}q^{n+1}$.\rq\rq {\em IEEE Trans. Inf. Theory,} vol. 58 no. 8, pp. 5534-5543, 2012.

\bibitem{Yong} Y. S. Kim, J. W. Jang, S. H. Kim, J. S. No, \lq\lq Linear complexity of quaternary sequences constructed from binary Legendre sequences.\rq\rq {\em ISITA 2012,} pp. 611-614, 2012.

\bibitem{Andrew Klapper} A. Klapper and  M. Goresky,  \lq\lq Feedback Shift Registers, 2-Adic Span, and Combiners with Memory.\rq\rq \emph{J. Cryptol.}, vol. 10, pp. 111-147, 1997.


\bibitem{Li Nian} N. Li and X. Tang, \lq\lq On the Linear Complexity of Binary sequences of Period $4N$ with Optimal Autocorrelation/Magnitude.\rq\rq {\em IEEE Trans. Inf. Theory,} vol. 57, no. 11, pp. 7597-7604, 2011.

\bibitem{Rainer} R. A. Rueppel, \lq\lq Linear Complexity and Random Sequences.\rq\rq {\em Advances in Cryptology-EUROCRYPT'85,} pp. 167-188, 1986.

\bibitem{Sun Yuhua-1} Y. Sun, Q. Wang, T. Yan, \lq\lq The exact autocorrelation distribution and 2-adic complexity of a class of binary sequences with almost optimal autocorrelation. \rq\rq {\em Cryptogr. Commun.,} Vol 10(3), pp. 467-477, 2018.

\bibitem{Sun Yuhua-2} Y. Sun, Q. Wang, T. Yan, \lq\lq A lower bound on the 2-adic complexity of the modified Jacobi sequence. \rq\rq {\em Cryptogr. Commun.,} Vol 11(2), pp. 337-349, 2019.

\bibitem{Sun Yuhua-3} Y. Sun, T. Yan, Z. Chen, L. Wang \lq\lq The 2-adic complexity of a class of binary sequences with optimal autocorrelation magnitude. \rq\rq {\em Cryptogr. Commun.,} https://doi.org/10.1007/s12095-019-00411-4.

\bibitem{Tian Tian} T. Tian and W. Qi,  \lq\lq 2-Adic Complexity of Binary $m$-Sequences.\rq\rq {\em IEEE Trans. Inf. Theory,}
vol. 56, no. 1, pp. 450-454, 2010.

\bibitem{Qi Wang} Q. Wang and X.  Du, \lq\lq The Linear Complexity of Binary sequences with Optimal Autocorrelation.\rq\rq {\em IEEE Trans. Inf. Theory,} vol. 56, no. 12, pp. 6388-6397, 2010.

\bibitem{Wang Qiu} Q. Wang, Y. Jiang, D. Lin, \lq\lq Linear complexity of binary generalized cyclotomic sequences over $\mathrm{GF}(q)$.\rq\rq {\em J. Complexity,} vol. 31, no. 5, pp. 731-740, 2015.


\bibitem{Zeng Xiangyong} Z. Xiao and X. Zeng, \lq\lq 2-Adic complexity of two classes of generalized cyclotomic binary sequences.\rq\rq Int. J. Found. Comput. S. 27, 879-893 (2016).


\bibitem{Xiong} H. Xiong, L. Qu, C. Li, S. Fu, \lq\lq Linear complexity of binary sequences with interleaved structure.\rq\rq {\em IET Communications,} vol. 7, no. 5, pp. 1688-1696, 2013.

\bibitem{Xiong Hai} H. Xiong, L. Qu, C. Li, \lq\lq A New Method to Compute the 2-Adic Complexity of Binary Sequences.\rq\rq {\em IEEE Trans. Inf. Theory,} vol. 60, no. 4, pp. 2399-2406, 2014.

\bibitem{Xiong Hai-2} H. Xiong, L. Qu, C. Li, \lq\lq 2-Adic complexity of binary sequences with interleaved structure.\rq\rq {\em Finite Fields Th. App.,} vol. 33, pp. 14-28, 2015.


\end{thebibliography}
\end{document}